\documentclass[epj]{svjour}
\pdfoutput=1

\usepackage{graphicx}
\usepackage{amsmath}

\begin{document}

\title{Unsupervised and semi-supervised clustering by message passing:\\ 
Soft-constraint affinity propagation}

\author{Michele Leone, Sumedha, and Martin Weigt}
%\affiliation{Institute for Scientific Interchange, Viale Settimio Severo 65, Villa Gualino, I-10133 Torino, Italy}
\institute{Institute for Scientific Interchange, Viale Settimio Severo 65, Villa Gualino, I-10133 Torino, Italy}

\date{\today}

\abstract{
Soft-constraint affinity propagation (SCAP) is a new statistical-physics 
based clustering technique \cite{SCAP}. First we give the derivation of a 
simplified version of the algorithm and discuss possibilities of 
time- and memory-efficient implementations. Later we give a detailed
analysis of the performance of SCAP on artificial data, showing that
the algorithm efficiently unveils clustered and hierarchical data
structures. We generalize the algorithm to the problem of semi-supervised
clustering, where data are already partially labeled, and clustering
assigns labels to previously unlabeled points. SCAP uses both the
geometrical organization of the data and the available labels assigned 
to few points in a computationally efficient way, as is shown on 
artificial and biological benchmark data.}


\authorrunning{M. Leone et al.}
\titlerunning{Soft-constraint affinity propagation}

\maketitle

\section{Introduction}

Clustering is a very important problem in  data analysis 
\cite{JAIN,DUDA}. Starting from a set of data points, one tries to group data 
such that points in one cluster are more similar in between each other
than points in different clusters. The hope is that such a grouping unveils
common functional characteristics. As an example, one of the currently most
important application fields for clustering is the informatical analysis
of biological high-throughput data, as given e.g. by gene expression data. 
Different cell states result in different expression patterns.

If data are organized in a well-separated way, one can use one of the many 
unsupervised clustering methods to divide them into classes \cite{JAIN,DUDA}; but
if clusters overlap at their borders or if they have involved shapes, these 
algorithms in general face problems.  However, clustering can still be achieved 
using a small fraction of previously labeled data (training set), making the 
clustering {\it semi-supervised} \cite{BOOK,DOMANY2}. While designing algorithms for 
semi-su\-per\-vised clustering, one has to be careful: They should efficiently
use both types of information provided by the geometrical organization of the
data points as well as the already assigned labels. 

In general there is not only one possible clustering. If one goes to a very
fine scale, each single data point can be considered its own cluster. On a
very rough scale, the whole data set becomes a single cluster. These two
extreme cases may be connected by a full hierarchy of cluster-merging events.

This idea is the basis of the oldest clustering method, which still is amongst 
the most popular one: {\it hierarchical agglomerative clustering} \cite{SOKAL,JOHNSON}. 
It starts with clusters being isolated points, and in each algorithmic step the two 
closest clusters are merged (with the cluster distance given, e.g., by the 
minimal distance between pairs of cluster elements), until only one big cluster
appears. This process can be visualized by the so-called dendrogram, which
shows clearly possible hierarchical structures. The strong point of this 
algorithm is its conceptual clarity connected to an easy numerical implementation.
Its major problem is that it is a greedy and local algorithm, no decision can
be reversed.

A second traditional and broadly used clustering method is {\it K-means 
clustering} \cite{MCQUEEN}. In this algorithm, one starts with a random assignment of
data points to $K$ clusters, calculates the center of mass of each cluster, 
reassigns points to the closest cluster center, recalculates cluster centers etc.,
until the cluster assignment is converged. This method is a very efficiently
implementable method, but it shows a strong dependence on the initial condition,
getting trapped by local optima. So the algorithm has to be rerun many times
to produce reliable clusterings, and the algorithmic efficiency is decreased.
Further on $K$-means clustering assumes spherical clusters, elongated clusters
tend to be divided artificially in sub-clusters.

A first statistical-physics based method is {\it super-para\-magnetic clustering}
\cite{DOMANY1,DOMANY2}.
The idea is the following: First the network of pairwise similarities becomes
preprocessed, only links to the closest neighbors are kept. On this sparsified
network a ferromagnetic Potts model is defined. In between the paramagnetic
high-temperature and the ferromagnetic low-temperature phase a super-paramagnetic
phase can be found, where already large clusters tend to be aligned. Using
Monte-Carlo simulations, one measures the pairwise probability for any two points
to take the same value of their Potts variables. If this probability is large
enough, these points are identified to be in the same cluster. This algorithm
is very elegant since it does not assume any cluster number of structure, nor uses
greedy methods. Due to the slow equilibration dynamics in the super-paramagnetic
regime it needs, however, the implementation of sophisticated cluster Monte-Carlo 
algorithms. Note that also super-paramagnetic clusterings can be obtained by
message passing techniques, but these require an explicit breaking of the 
symmetry between the values of the Potts variables to give non-trivial results.

Also in the last years, many new clustering methods are being proposed. One
particularly elegant and powerful method is {\it affinity propagation} (AP) 
\cite{FREY}, which gave also the inspiration to our algorithm. The approach is 
slightly different: Each data point has to select an exemplar in between all 
other data points. This shall be done in a way to maximize the overall similarity
between data points and exemplars. The selection is, however, restricted by a 
hard constraint: Whenever a point is chosen as an exemplar by somebody else,
it is forced to be also its own self-exemplar. Clusters are consequently
given as all points with a common exemplar. The number of clusters is regulated
by a chemical potential (given in form of a self-similarity of data points),
and good clusterings are identified via their robustness with respect to
changes in this chemical potential. The computational hard task to optimize
the overall similarity under the hard constraints is solved via message
passing \cite{YEDIDIA,MAXSUM}, more precisely via belief propagation, which
are equivalent to the Bethe-Peierls approximation / the cavity
method in statistical physics \cite{MezardParisi,MyBook}. Despite the very good
performance on test data, also AP has some drawbacks: It assumes again
more or less spherical clusters, which can be characterized by a single
cluster exemplar. It does not allow for higher order pointing processes.
A last concern is the robustness: Due to the hard constraint, the change
of one single exemplar may result in a large avalanche of other changes.

The aim of {\it soft-constraint affinity propagation} (SCAP) is to use the
strong points and ideas of affinity propagation -- the exemplar choice 
fulfilling a global optimization principle, the computationally efficient 
implementation via message-passing techniques -- but curing the problems
arising from the hard constraints. In \cite{SCAP} we have 
proposed a first version of this algorithm, and have shown that on 
gene-expression data it is very powerful. In this article, we propose a 
simplified version which is more efficient. Finally we show that SCAP also 
allows for a particularly elegant generalization to the semi-supervised case, 
{\it i.e.}~to the inclusion of partially labeled data.
As shown in some artificial and biological benchmark data, the partial labeling 
allows to extract the correct clustering even in cases where the unsupervised
algorithm fails. 

The plan of the paper is the following: After this Introduction, we present
in Sec.~\ref{sec:scap} the clustering problem and the derivation of SCAP, and
we discuss time- and memory-efficient implementations which become important
in the case of huge data sets. In Sec.~\ref{sec:data} we test the performance
of SCAP on artificial data with clustered and hierarchical structures. 
Sec.~\ref{sec:semi} is dedicated to the generalization to semi-supervised
clustering, and we conclude in the final Sec.~\ref{sec:conclusion}.

\section{The algorithm}
\label{sec:scap}

\subsection{Formulation of the problem}

The basic input to SCAP are pairwise similarities $S(\mu,\nu)$ between
any two data points $\mu,\nu\in \{1,...,N\}$. In many cases, these
similarities are given by the negative (squared) Euclidean distances
between data points or by some correlation measure (as Pearson
correlations) between data points. In principle they need not even to
be symmetric in $\mu$ and $\nu$, as they might represent conditional
dependencies between data points. The choice of the correct similarity
measure will for sure influence the quality and the details of the
clusterings found by SCAP, it depends on the nature of the data
which shall be clustered. Here we assume therefore the similarities to
be given.

The main idea of SCAP is that each data point $\mu$ selects some other
data point $\nu$ as its {\it exemplar}, i.e. as some reference point
for itself. The exemplar choice is therefore given by a mapping
\begin{equation}
  \label{eq:c_map}
  {\mathbf c}:\ \ \{1,...,N\} \ \mapsto\  \{1,...,N\} 
\end{equation}
where, in difference to the original AP and the previous version of
SCAP, no self-exemplars are allowed:
%\footnote{Compared to SCAP as
%presented in \cite{SCAP}, this corresponds to the the
%case where self-similarities $S(\mu,\mu)$ are sent to
%$-\infty$. Practically, this was already done in the analysis of
%\cite{SCAP}, even if there the direct simplifications in
%the algorithm were not taken into account.}
\begin{equation}
  \label{eq:no_self_exemplar}
  \forall \mu\in\{1,...,N\}:\ \ c_\mu \neq \mu\ .
\end{equation}
The mapping ${\mathbf c}$ defines a directed graph with links going from data
points to their exemplars, and clusters in this approach correspond to
the connected components of (an undirected version) this graph.

The aim in constructing ${\mathbf c}$ is to minimize the Hamiltonian, or cost
function,
\begin{equation}
  \label{eq:H} 
  {\cal H} ({\mathbf c}) = - \sum_{\mu=1}^N S(\mu,c_\mu)\ +\ 
  p\ {\cal N}_c\ ,
\end{equation}
with ${\cal N}_c$ being the number of distinct selected exemplars.
This Hamiltonian consists of two parts: The first one is the negative
sum of the similarities of all data points to their exemplars, so the
algorithm tries to maximize this accumulated similarity. However, this
term alone would lead to a local greedy clustering where each data
point chooses its closest neighbor as an exemplar. The resulting
clustering would contain ${\cal O}(N)$ clusters, so increasing the
amount of data would lead to more instead of better defined clusters.
The second term serves to {\it compactify} the clusters: $\chi_\mu$ is
one iff $\mu$ is an exemplar, so each exemplar has to pay a {\it
penalty} $p$. Since this penalty does not depend on how many data
points actually choose $\mu$ as their exemplar (the in-degree of
$\mu$), mappings ${\bf c}$ with few exemplars of high in-degree are
favored, leading to more compact clusters. In this way, the parameter
$p$ controls the cluster number, robust clusterings are recognized due
to their stability under changing $p$. Since the cluster number is not
fixed a priori, SCAP also recognizes successfully a hierarchical
cluster organization.

For later convenience we express the exemplar number as
\begin{equation}
  {\cal N}_c = \sum_{\mu=1}^N \chi_\mu({\mathbf c})\ ,
\end{equation}
using an indicator function
\begin{equation}
  \chi_\mu({\mathbf c}) = \left\{
  \begin{array}{ccl}
    1 && {\rm if}\ \exists \nu:\ c_\nu=\mu\\
    0 && {\rm else}
  \end{array}
  \right.
\end{equation}
which denotes the {\it soft local constraint} acting on each data point.
 
Note that this problem setting is slightly different from the one used in 
the first derivation of SCAP in \cite{SCAP}. There self-exemplars
were allowed, and only selecting an exemplar which was not a self-exemplar
led to the application of the penalty $ p$. The number of self-exemplars 
itself was coupled to a second parameter, the self-similarity. In 
\cite{SCAP} we already found that the best results were obtained
for very small self-similarities. Actually the algorithm presented here can
be obtained from the previous formulation by explicitly sending all 
self-similarities $S(\mu,\mu)\to -\infty$. The resulting formulation is
easier both in implementation and interpretation since it does not include
self-messages.

\subsection{Derivation of the algorithm}

The exact minimization of this Hamiltonian is a computationally hard problem: 
There are $(N-1)^N$ possible configurations $c$ to be tested, resulting in a 
potentially super-exponential running time of any exact algorithm. We therefore
need efficient heuristic approaches which, even if not guaranteeing to find
the true optimum, are algorithmically feasible. 

An approach related to the statistical physics of disordered systems is the
implementation of message-passing techniques, more precisely of the belief
propagation algorithm \cite{YEDIDIA,MAXSUM}. The latter is equivalent to an 
algorithmic interpretation of the Bethe-Peierls approximation in statistical physics: 
Instead of solving exactly the thermodynamics of the problem, we use a refined 
mean-field method.

To do so, we first introduce a formal inverse temperature $\beta$ and the
corresponding Gibbs weight
\begin{equation}
  \label{eq:gibbs}
  {\mathbf P} ({\mathbf c}) \sim \exp\{-\beta {\cal H}({\mathbf c}) \}\ .
\end{equation}
The temperature will be sent to zero at the end of the calculations, to obtain
a weight concentrated completely in the ground states of ${\cal H}$. In 
principle one should optimize ${\mathbf P}({\mathbf c})$ with respect to the joint 
choice of all exemplars, we will replace this by the independent optimization of all
marginal single-variable probabilities. We thus need to estimate the probabilities
\begin{equation}
\label{eq:marginal}
P_\mu(c_\mu) = \sum_{\{c_\nu;\nu\neq\mu\}} {\mathbf P} ({\mathbf c})
\end{equation}
which in principle contain a sum over the $(N-1)^{N-1}$ configurations of all 
other variables. From this marginal probability we can define an exemplar choice
as
\begin{equation}
  \label{eq:argmax}
  c_\mu^\star = \underset{c_\mu}{\rm argmax} \lim_{\beta\to\infty} P_\mu(c_\mu)\ .
\end{equation}
Note that this becomes the correct global minimum of ${\mathbf P}$ if the 
latter is non-degenerate which is a reasonable assumption in the case of 
real-valued similarities $S(\mu,\nu)$.

\begin{figure}[htb]
%\vspace{0.2cm}
\begin{center}
\includegraphics[width=\columnwidth]{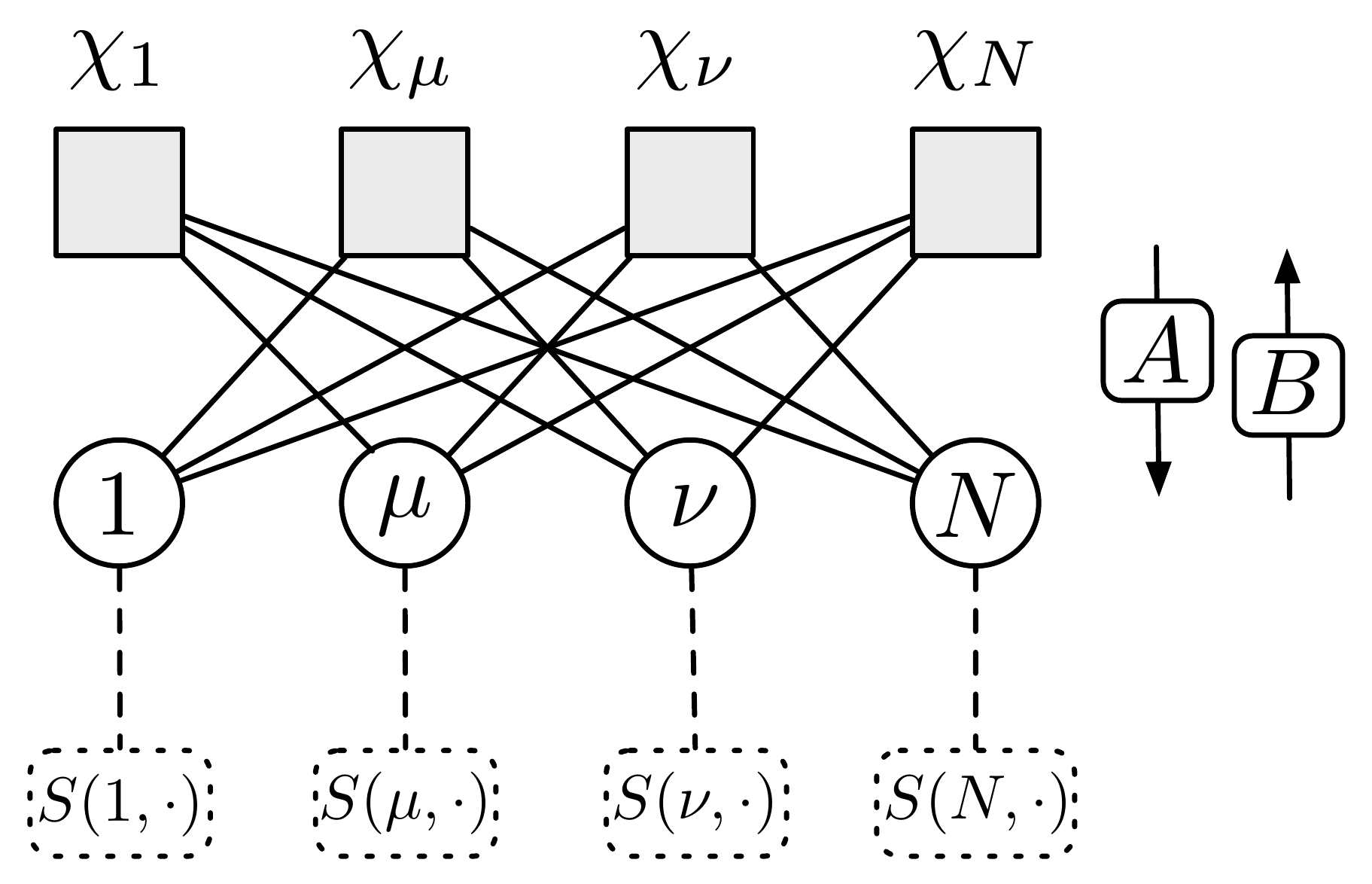}
\end{center}
\caption{Factor graph for SCAP: Circles denote variable nodes, related to the
variables $c_\mu$, whereas squares denote the constraints $\chi_\mu$. A link is
drawn whenever a variable compares in a constraint, i.e. all variable nodes
$\nu\neq\mu$ are connected to factor node $\chi_\mu$. Similarities act as
external $(N-1)$-dimensional fields on the variables. The figure also displays
the two message types send from variables to constraints and back.}
\label{fig:factor_graph_scap}
\end{figure}

We want to estimate these marginal distributions using belief propagation, or
equivalently the Bethe-Peierls approximation. For doing so, we first represent
the problem by its factor graph as given in Fig.~\ref{fig:factor_graph_scap}.
The variables are represented by circular variable nodes, the constraints 
$\chi_\mu$ by square factor nodes. Due to the special structure of the
problem, every variable node corresponds to exactly one factor node. Each
factor node is connected to all variable nodes which are contained in the
constraint (which are all but the one corresponding to the factor node).
The similarities act locally on variable nodes, they can be interpreted as 
$(N-1)$-dimensional local vector fields. 

Belief propagation works via the exchange of messages between variable and factor
nodes. Let us denote first $A_{\mu\to\nu}(c_\nu)$ the message sent from constraint $\mu$ to
variable $\nu$, measuring the probability that $\mu$ forces $\nu$ to select $c_\nu$
as its exemplar. Second we introduce $B_{\nu\to\mu}(c_\nu)$ as the probability that
variable $\nu$ would choose $c_\nu$ as its exemplar without the presence of constraint
$\mu$. Than we can write down closed iterative equations, called belief-propagation
equations,
\begin{eqnarray}
\label{eq:BP}
A_{\mu\to\nu}(c_\nu) & \propto & \prod_{\lambda\neq\mu,\nu} 
\left[\sum_{c_\lambda} B_{\lambda\to\mu}(c_\lambda) \right]
\exp\{-\beta\, p\, \chi_\mu({\mathbf c})\} \nonumber\\
B_{\mu\to\nu}(c_\mu) & \propto & \prod_{\lambda\neq\mu,\nu}
A_{\lambda\to\mu}(c_\mu) \exp\{ \beta\, S(\mu,c_\mu)\} \ .
\end{eqnarray}
We see that the message $A_{\mu\to\nu}$ from constraint $\mu$ to variable $\nu$ 
depends on the choices all other variables would take without constraint $\mu$,
times the Gibbs weight of constraint $\chi_\mu$. The message $B_{\mu\to\nu}$
from variable $\mu$ to constraint $\nu$ depends on the messages from all other
constraints to $\mu$, and the local field $\vec S$ on $\mu$. The approximate 
character of belief propagation stems from the fact that the joint distributions 
over all neighboring variables is taken to be factorized into single variable
quantities. Having solved these equations we can easily estimate the true 
marginal distributions
\begin{equation}
\label{eq:p_mu}
P_{\mu}(c_\mu) \propto  \prod_{\lambda\neq\mu}
A_{\lambda\to\mu}(c_\mu) \exp\{ \beta\, S(\mu,c_\mu)\}
\end{equation}
which are the central quantities we are looking for.

However, looking at the first of Eqs.~(\ref{eq:BP}), we realize that it still
contains the super-exponential sum. Further on, we need a memory space of 
${\cal O}(N^3)$ to store all these messages, which is practical only for small
and intermediate data sets. This problem can be resolved exactly by
realizing that $A_{\mu\to\nu}(c_\mu)$ takes only two values for fixed $\mu$ and
$\nu$, namely $A_{\mu\to\nu}(\mu)$ and $A_{\mu\to\nu}(c\neq\mu)$
\footnote{This observation was first done in the case of original
AP in \cite{FREY}, and can be simply extended to our model}. We therefore 
introduce the reduced messages
\begin{eqnarray}
\tilde A_{\mu\to\nu} &=& \frac{A_{\mu\to\nu}(\mu)}{A_{\mu\to\nu}(c\neq\mu)}
\nonumber\\
\tilde B_{\mu\to\nu} &=& B_{\mu\to\nu}(\nu)\ .
\end{eqnarray}
After a little book-keeping work to consider all possible cases, the sums
in Eqs.~(\ref{eq:BP},\ref{eq:p_mu}) can be performed analytically resulting 
in a set of equivalent relations
\begin{eqnarray}
\label{eq:scap_T}
\tilde A_{\mu\to\nu} & =& \left[ 1+(e^{\beta p}-1) 
\prod_{\lambda\neq\mu,\nu} (1-\tilde B_{\lambda\to\mu}) 
\right]^{-1} \nonumber\\
\tilde B_{\mu\to\nu} & =& \left[ 1+ \sum_{\lambda\neq\mu,\nu} 
e^{ \beta S(\mu,\lambda)-\beta S(\mu,\nu) } \tilde A_{\lambda\to\mu} \right]^{-1}
\nonumber\\
P_\mu(c) & =& \frac{e^{\beta\,S(\mu,c)}  \tilde A_{c\to\mu}}
{\sum_{\lambda\neq\mu}e^{\beta\,S(\mu,\lambda)}  \tilde A_{\lambda\to\mu}}\ .
\end{eqnarray}
These equations are the {\it finite-temperature SCAP equations}. Note that the 
complexity of evaluating the first line is decreased from ${\cal O}(N^N)$ to
${\cal O}(N)$ and therefore feasible even for very large data sets. Also the 
memory requirements are decreased to ${\cal O}(N^2)$. As we will see later on,
a clever implementation will, in particular in the zero-temperature limit,
further decrease time- and space-com\-ple\-xity.

\subsection{SCAP in the zero-temperature limit}

Even if Eqs.~(\ref{eq:scap_T}) are already relatively simple, the zero temperature
limit of these equations becomes even simpler and bears a very intuitive 
interpretation. To achieve this limit, we have to transform the variables in
the equations from probabilities to local fields, and introduce
\begin{eqnarray}
a_{\mu\to\nu} &=& \frac 1\beta \ln \tilde A_{\mu\to\nu} \nonumber\\ 
r_{\mu\to\nu} &=& \frac 1\beta \ln \frac {\tilde B_{\mu\to\nu}}
{1-\tilde B_{\mu\to\nu}}\ .
\end{eqnarray}
We call $a_{\mu\to\nu}$ the {\it availability} of $\mu$ to be an exemplar for
$\nu$, whereas $r_{\mu\to\nu}$ measures the {\it request} of $\mu$ to point
$\nu$ to be its exemplar. Using the fact that sums over various exponential terms 
in $\beta$ are dominated by the maximum term, we readily conclude
\begin{eqnarray}
\label{eq:SCAP1}
r_{\mu\to\nu} &=&
S(\mu,\nu) -  {\rm max}_{\lambda\neq\mu,\nu}
\left[ S(\mu,\lambda) + a_{\lambda\to\mu} \right] \nonumber\\
a_{\mu\to\nu} &=& {\rm min} \left[ 0,\, - p
+ \sum_{\lambda\neq\mu,\nu} {\rm max}
(0,\, r_{\lambda\to\mu}) \right] 
\end{eqnarray}
to hold for these two fields.

These equations have a very nice and intuitive interpretation in terms of a 
social dynamics of exemplar selection. The system tries to maximize its overall
similarity (or gain) which is the sum over all similarities between data points
and their exemplars, but each exemplar has to pay a penalty $ p$. Therefore
each data point $\mu$ sends requests to all their neighbors $\nu$, which are 
composed by two contributions: The similarity to the neighbor itself, minus the 
maximum over all similarities to the other points $\lambda\neq\mu,\nu$ - the latter 
already being corrected for by the availability of the other points to be an 
exemplar. Now, data points $\mu$ communicate their availability to be an exemplar 
for any other data point $\nu$. For doing so, they sum up all positive requests 
from further points $\lambda\neq\mu,\nu$, and compare it to the penalty they have to
pay in case they accept to be an exemplar. If the accumulated positive requests are
bigger than the penalty, $\mu$ agrees right away to be the exemplar for $\nu$. If on 
the other hand the penalty is larger than the requests, $\mu$ communicates to
$\nu$ the difference - so the answer is not a simple ``no'' but is weighed. Point
$\nu$ should overcome this difference with its similarity.

Consequently the exemplar choice of $\mu$ happens via the selection of the 
neighbor $\nu$ who has the highest value of the similarity corrected by the
availability of $\nu$ for $\mu$, i.e. we have
\begin{equation}
\label{eq:SCAP2}
c_\mu^\star = \underset{\nu}{\rm argmax} \left[  S(\mu,\nu)+a_{\nu\to\mu}
\right]\ .
\end{equation}
Eqs.~(\ref{eq:SCAP1},\ref{eq:SCAP2}) are called {\it soft-constraint affinity
propagation}. They can be solved by first iteratively solving (\ref{eq:SCAP1}),
and then plugging the solution into (\ref{eq:SCAP2}). The next two sub-sections 
will show how this can be done in a time- and memory efficient way.

\subsection{Time-efficient implementation}

The iterative solution of Eqs.~(\ref{eq:SCAP1}) can be implemented in the 
following way:
\begin{enumerate}
\item Define the similarity $S(\mu,\nu)$ for  each set of data points. Choose 
the values of the self-similarity $\sigma$ and of the constraint strength $ p$. 
Initialize all $a(\mu,\nu)=r(\mu,\nu)=0$ 
\item For all $\mu\in\{1,...,N\}$, first update the $N$ {\it requests}
$r_{\mu\to\nu}$ and then the $N$ {\it availabilities} $a_{\mu\to\nu}$, using 
Eqs.~(\ref{eq:SCAP1}).
\item Identify the exemplars $c_\mu^\star$ by looking at the maximum value of
$S(\mu,\nu)+a_{\nu\to\mu}$ for given $\mu$, according to Eq.~(\ref{eq:SCAP2}).
\item Repeat steps 2-3 till there is no change in exemplars 
for a large number of iterations (we used 10-100 iterations). If not
converged after $T_{max}$ iterations (typically 100-1000), stop the
algorithm.
\end{enumerate}

Three notes are necessary at this point:
\begin{itemize}
\item Step 3 is formulated as a sequential update: For each data point $\mu$, 
all outgoing responsibilities and then all incoming availabilities are updated 
before moving to the next data point. In numerical experiments this was found
to converge faster and in a larger parameter range than the damped parallel
update suggested by Frey and Dueck in \cite{FREY}. The actual implementation
uses a random sequential update, i.e. each time step 3 is performed, we generate
a random permutation of the order of  the  $\mu\in\{1,...,N\}$.
\item The naive implementation of the update equations (\ref{eq:SCAP1}) requires 
${\cal O}(N^2)$ updates, each one of computational complexity ${\cal O}(N)$. 
A factor $N$ can be gained by first computing the unrestricted max and sum 
once for a given $\mu$, and then implying the restriction only inside the 
internal loop over $\nu$. Like this, the total complexity of a global update is 
${\cal O}(N^2)$ and thus feasible even for very large data sets.
\item Belief propagation on loopy graphs is not guaranteed to converge. We observe,
that even in cases where the messages do not converge to a fixed point but go on
fluctuating, the exemplar choice converges. In our algorithm, we therefore
apply frequently the stationarity of ${\mathbf c}^\star$ as a weaker convergence 
criterion than message convergence.
\end{itemize}
 
\subsection{Memory-efficient implementation}

Another problem of SCAP can be its memory size, Eqs.~(\ref{eq:SCAP1}) require the
storage of three arrays of size $N^2$. This can be a problem if we consider very
large data sets. A particularly important example are gene-expression data, which
may contain more than 30,000 genes. If one wants to cluster these genes to identify
coexpressed gene groups, the required memory size becomes fastly much larger than
the working memory of a standard desktop computer, restricting the size of data
sets to approximately $N<10^4$.

However, this problem can be resolved in the zero-temperature equations by not 
storing messages and similarities (which are indexed by two numbers) but only
site quantities (which are indexed by a single number) reducing thus the
memory requirements to ${\cal O}(N)$. This allows to treat even the largest
available data sets efficiently with SCAP.

As a first step, we note that in most cases data are multi-dimensional. For example
in gene expression data, a typical data sets contains about 100 
micro-arrays measuring simultaneously 5,000-30,000 genes. If we want to cluster arrays,
for sure a direct implementation of Eqs.~(\ref{eq:SCAP1},\ref{eq:SCAP2}) is best.
In particular only the similarities are needed actively instead of the initial
data points. If, on the other hand, we want to clusterize genes, it is more 
efficient to calculate similarities whenever needed from the original data,
instead of memorizing the huge similarity matrix.

Once this is implemented, we can also get rid of the messages $a_{\mu\to\nu}$ and
$r_{\mu\to\nu}$. First we introduce
\begin{eqnarray}
h_\mu^{(1)} &=& 
\underset{\lambda\neq\mu}{\rm max} \left[ S(\mu,\lambda) + a_{\lambda\to\mu} \right]
\nonumber\\
c_\mu^{(1)} &=& 
\underset{\lambda\neq\mu}{\rm argmax} \left[ S(\mu,\lambda) + a_{\lambda\to\mu} \right]
\nonumber\\
h_\mu^{(2)} &=& 
\underset{\lambda\neq\mu,c_\mu^{(1)}}{\rm max} \left[ S(\mu,\lambda) 
+ a_{\lambda\to\mu} \right]\ .
\end{eqnarray}
These quantities, together with the similarities (directly calculated from the 
original data) are sufficient to express all requests,
\begin{equation}
r_{\mu\to\nu} = S(\mu,\nu) - h_\mu^{(1)} + \left(h_\mu^{(1)}-h_\mu^{(2)}\right)
\, \delta_{\nu,c_\mu^{(1)}}
\end{equation}
with $\delta_{\cdot,\cdot}$ being the Kronecker-symbol. A similar step can be done
for the availabilities. We introduce
\begin{equation}
u_\mu =  \sum_{\lambda\neq\mu} {\rm max} (0,\, r_{\lambda\to\mu})
\end{equation}
and express the availability as
\begin{eqnarray}
a_{\mu\to\nu} &=& {\rm min} \Big[ 0,\, - p + u_\mu  \\ & & \left.
- {\rm max}\left\{0,
\, S(\nu,\mu) - h_\nu^{(1)} + \left(h_\nu^{(1)}-h_\nu^{(2)}\right)
\, \delta_{\mu,c_\nu^{(1)}} \right\} \right] \nonumber
\end{eqnarray}

Note that after convergence we have trivially
\begin{equation}
c_\mu^\star = c_\mu^{(1)} 
\end{equation}
for all $\mu\in\{1,...,N\}$.

In this way, instead of storing $S(\mu,\nu)$, $a_{\mu\to\nu}$ and $r_{\mu\to\nu}$
we have to store only the data, $h_\mu^{(1,2)},\ c_\mu^{(1)}$ and $u_\mu$. The 
largest array is the data set itself, all
other memorized quantities require much less size. For large data sets, in this way
the memory usage becomes much more efficient. Even if the algorithm requires more
steps to be executed (similarities and messages have to be computed whenever they are
needed, instead of a single time in each update step), the more efficient memory usage
leads to strongly decreased running times.

\section{Artificial data}
\label{sec:data}

In \cite{SCAP} we have shown that SCAP is able to successfully cluster 
biological data coming from gene-expression arrays. This is true also for the simplified
version derived in the present work. Here we aim, however, at a more theoretical analysis
on artificial data which will bring light into some characteristics of SCAP, and which
will allow for a more detailed comparison to the performance of AP as defined
originally in \cite{FREY}. To start with, we first consider numerically data
having only one level of clustering, later on we extend this study to more than one
level of clusters, i.e. to a situation where clusters of data points itself are organized
in larger clusters.

\subsection{One cluster level}

\begin{figure}[htb]
%\vspace{0.2cm}
\begin{center}
\includegraphics[width=\columnwidth]{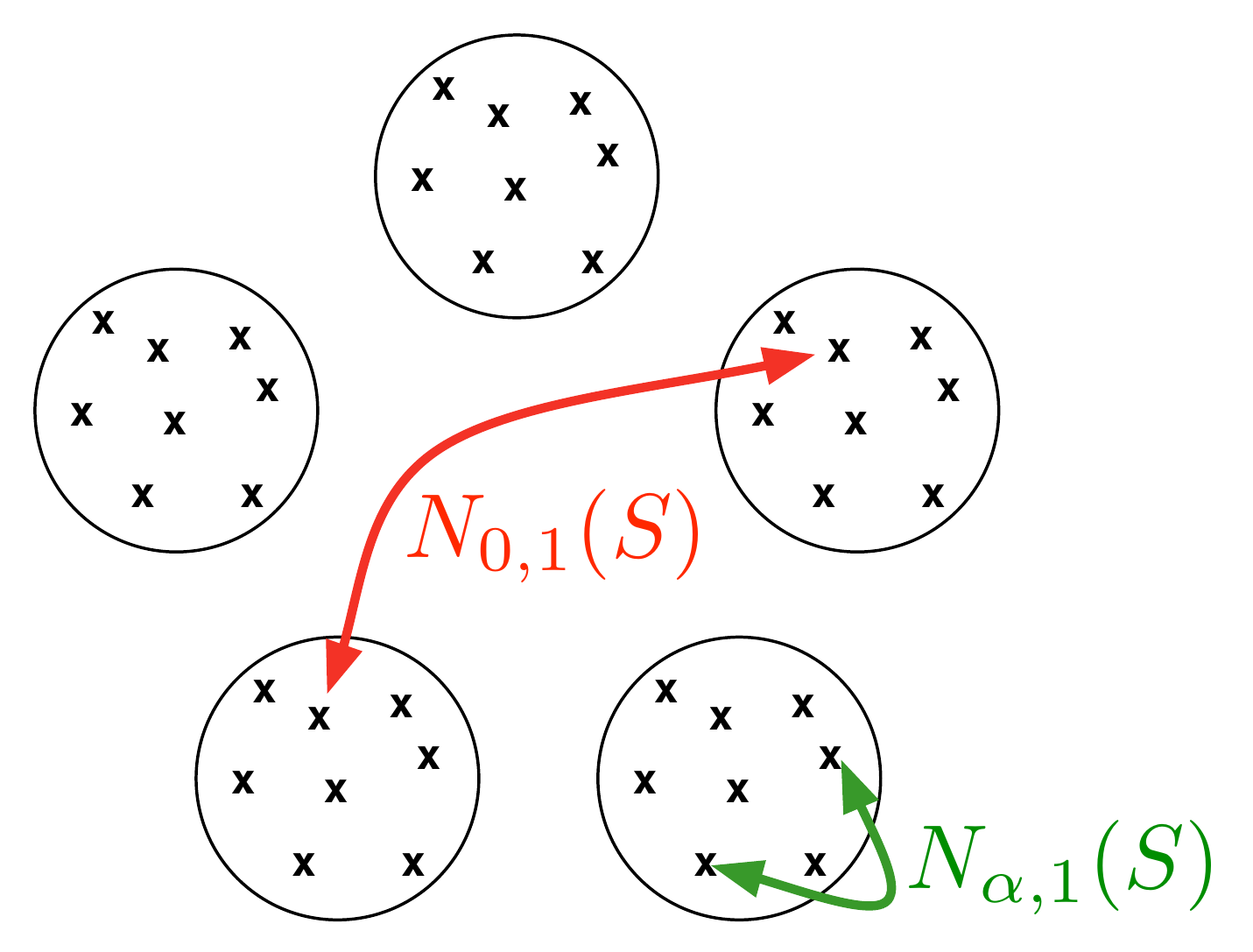}
\end{center}
\caption{Artificial data set for testing SCAP: $N$ data points
(crosses) are organized into $q$ clusters (full circles), similarities
for pair of points in the same cluster are drawn independently from a
Gaussian $N_{\alpha,1}(S)$ of mean $\alpha$ and variance 1, between
clusters from $N_{0,1}(S)$. The parameter $\alpha>0$ determines the
separability of the clusters.}
\label{fig:clusters}
\end{figure}

The first step is very simple: We define an artificial data set
having only one level of clustering. We therefore start with $N$ data
points which are divided into $q$ equally sized subsets. For each pair
inside such a subset we draw randomly and independently a similarity
from a Gaussian of mean $\alpha$ and variance one, whereas pair
similarities of data points in different clusters are drawn as
independent Gaussian numbers of zero mean and variance one,
cf. Fig.~\ref{fig:clusters} for an illustration. The parameter
$\alpha$ controls the separability of the clusters, for small
$\alpha<1$ clusters are highly overlapping, and SCAP is expected to be
unable to separate the $q$ subsets, whereas for large values $\alpha >
3$ a good separability is expected. Alternative definitions of the
similarities where data points are defined via high-dimensional data
with higher intra-cluster correlations, lead to similar results and
are not discussed here.

\begin{figure}[htb]
%\vspace{0.2cm}
\begin{center}
\includegraphics[width=\columnwidth]{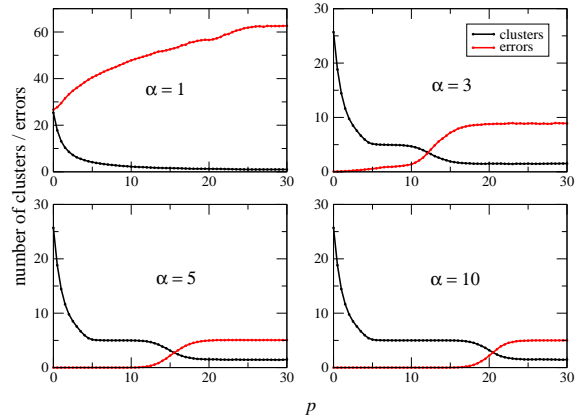}
\end{center}
\caption{Results of SCAP as a function of $ p$ for various
values of $\alpha$. Displayed are the number of clusters (black lines)
and errors (red lines). Results are for $N=100$, averaged over 1000
samples.}
\label{fig:scap_alpha}
\end{figure}

First we study the dependence of the SCAP results on the parameter
$\alpha$, see Fig.~\ref{fig:scap_alpha}. For $\alpha=1$, we see that
there is no signal at all at five clusters, and the error number
(measured as the number of points having exemplars in a different
cluster) grows starting from a high value. Data are completely mixed,
which is clear since $N_{0,1}$ and $N_{1,1}$ are strongly
overlapping. For $\alpha=3$, a clear plateau at five clusters appears,
and the error rate until this plateau is low. Only when we force the
system to form less than five clusters, the error rate starts to grow
considerably. This picture becomes even more pronounced for larger
$\alpha$; the distributions of intra- and inter-cluster
similarities are perfectly separated, SCAP makes basically no errors
until it is forced to do so since it forms less than five clusters.
The error rate is not found to go beyond five errors, which is very
small considering the fact that at least four errors are needed to 
interconnect the five clusters.

\begin{figure}[htb]
%\vspace{1cm}
\begin{center}
\includegraphics[width=\columnwidth]{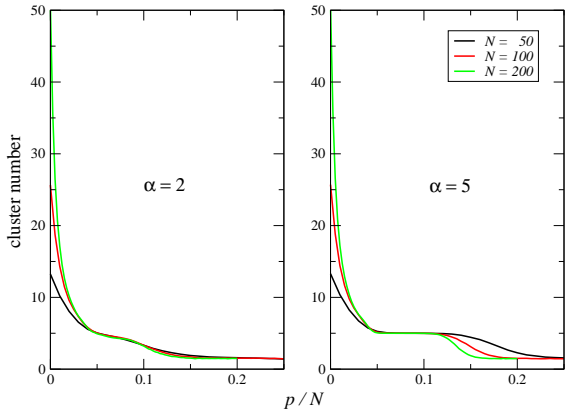}
\end{center}
\caption{Dependence of the SCAP results for different values of $N$. 
Curves result from averages over 1000 random samples.}
\label{fig:scap_N}
\end{figure}

Fig.~\ref{fig:scap_N} shows the $N$-dependence of the SCAP results.
The parameter $ p$ has to be rescaled by $N$ to re-balance the
increased number of contributions to the overall similarity in the
model's Hamiltonian. One sees that the initial cluster number for
$ p = 0$ is linear in $N$, but the penalty successfully forces
the system to show a collective behavior with macroscopic clusters.
The plateau length for different $N$ values is comparable, even if for
larger $N$ the decay from the plateau to 1-2 clusters is much more
abrupt.

\begin{figure}[htb]
%\vspace{1cm}
\begin{center}
\includegraphics[width=\columnwidth]{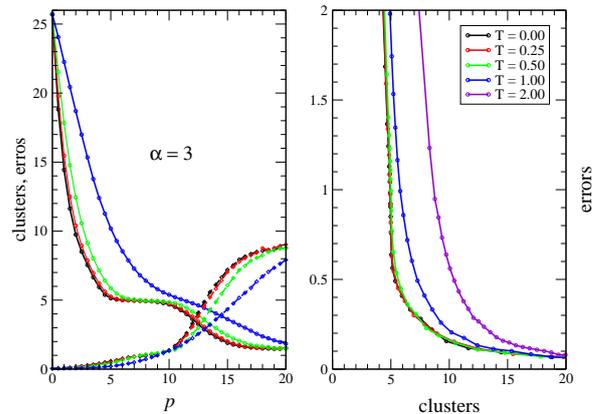}
\end{center}
\caption{Temperature dependence of the SCAP results for $N=100,\,
\alpha=3$.  The left figure shows the $ p$ dependence of the
cluster number (full lines) and of the error number (dashed
lines). The right figure shows a parametric plot of the numbers of
errors vs. clusters. Curves result from averages over 1000 random
samples.}
\label{fig:scap_T}
\end{figure}

Fig.~\ref{fig:scap_T} studies the influence of the formal temperature
on SCAP. In some cases finite-temperature SCAP shows more efficient
convergence, so it is interesting to see how much information is lost
by increasing the formal temperature. The left panel of
Fig.~\ref{fig:scap_T} represents again the cluster number (resp. error
number) as a function of $ p$. We see that for very low
temperature ($T=0.25$ in the example) results are hardly
distinguishable from the zero-temperature results. If we further
increase the temperature we observe that the plateau at five clusters
becomes less pronounced and shifted to larger $ p$. To get rid
of this shift, we show in the right panel a parametric plot of the two
most interesting quantities: The error number as a function of the
cluster number. This plot shows again that the errors start to grow
considerably (with decreasing cluster number) as soon as we go below
five clusters. For low enough temperatures, the curves practically
collapse, so very few of the clustering information is lost. Only for
higher temperatures the error number starts to grow already at higher
cluster numbers. The pronounced change when we cross the number of clusters
is lost. Therefore, as long as the plateau is pronounced in the left
panel, also the error number remains almost as low as in zero
temperature on the plateau.

\begin{figure}[htb]
%\vspace{1cm}
\begin{center}
\includegraphics[width=\columnwidth]{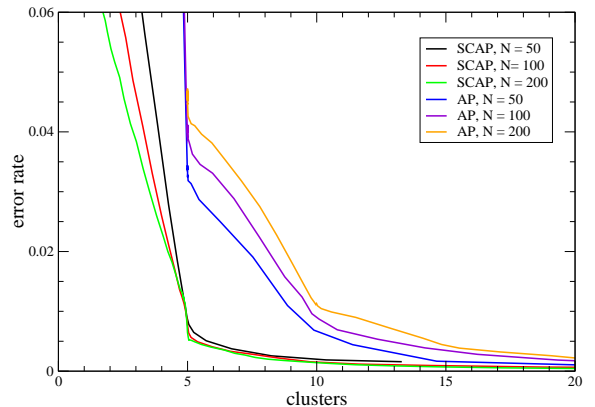}
\end{center}
\caption{SCAP vs. AP: The number of errors (divided by $N$) is plotted
against the cluster number, for $\alpha=3$ and various values of
$N$. Curves result from averages over 1000 samples.}
\label{fig:scap_ap}
\end{figure}

Last but not least, we compare the performance of SCAP to the original
AP proposed in \cite{FREY}. AP shows a slightly different
behavior than SCAP. The latter has only one plateau at the correct
cluster number, whereas AP shows a long plateau at five clusters, but
also less pronounced shoulders at multiples of this number. Both
algorithms can be compared directly when plotting the number of errors
against the cluster number, see Fig.~\ref{fig:scap_ap}. Note that in
principle this test is a bit easier for AP since a part of the data
points are self-exemplars, which are not counted as
errors. Nevertheless SCAP shows much less errors, in particular also
on the plateau of five clusters. The hard constraint in AP forbidding
higher order pointing processes is too strong even for a simple data
set as the one considered here, simply because the random generation
of the similarities makes all points on statistically equivalent, not
preferring one as a cluster center. The more flexible structure of
SCAP is able to cope with this fact and is therefore results in a
more precise clustering. Note that this difference increases with
growing size $N$ of the data set: Whereas the error number of SCAP at
five clusters slightly decreases with $N$, the corresponding number
for AP grows. This is again due to the hard constraint which forces
inside a cluster more and more data points to refer to the cluster
exemplar.

\subsection{Hierarchical cluster organization}

\begin{figure}[htb]
%\vspace{0.2cm}
\begin{center}
\includegraphics[width=\columnwidth]{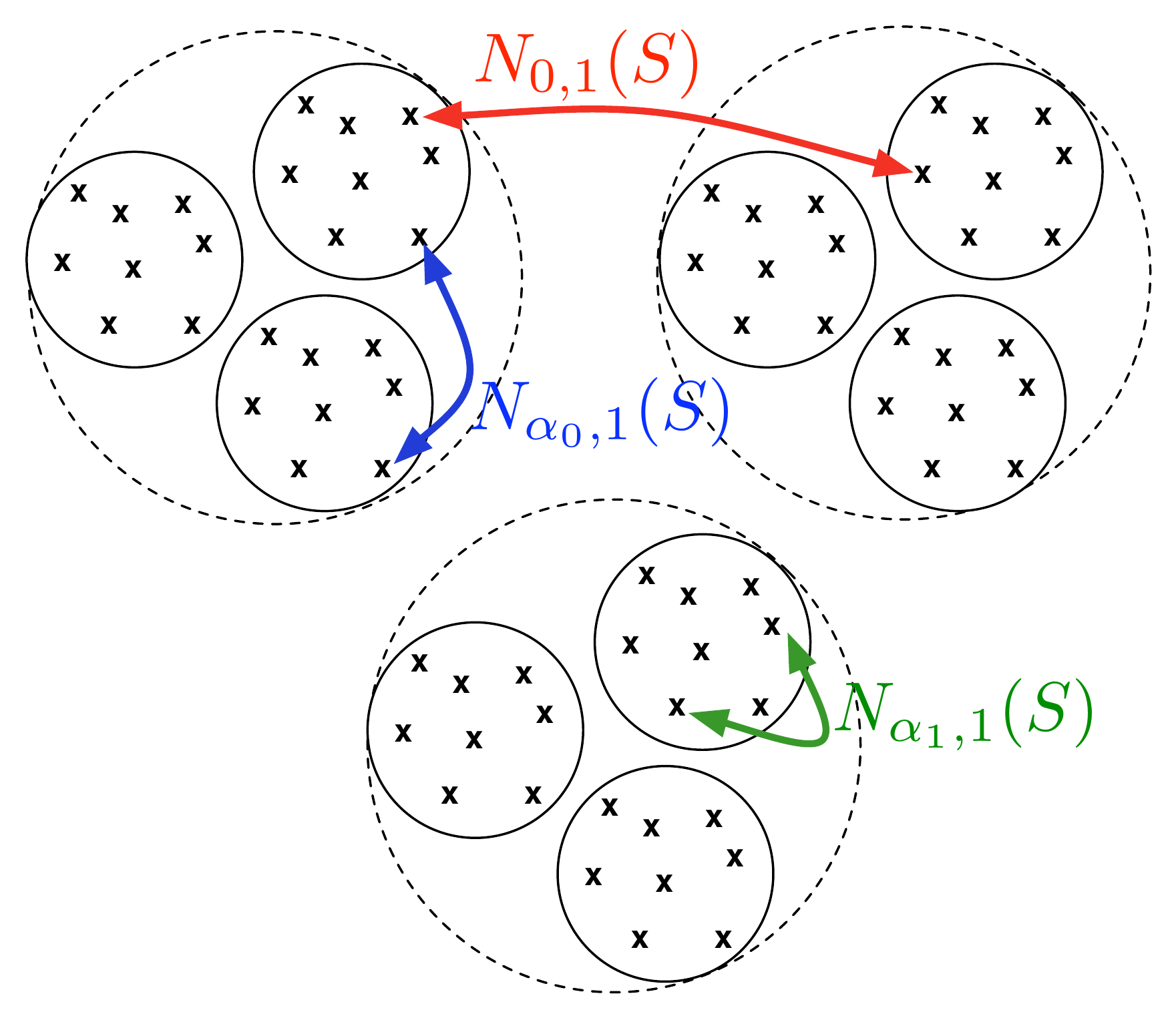}
\end{center}
\caption{Artificial data with two-level hierarchical
organization. Data (crosses) are organized in clusters (full circles),
which themselves are collected in larger clusters (dashed
circles). Similarities are drawn from Gaussians as shown in the
figure, with $0<\alpha_0<\alpha_1$.}
\label{fig:clusters_2level}
\end{figure}

To test if SCAP is also able to detect a hierarchical cluster
organization we have slightly modified the generator, as shown in
Fig.~\ref{fig:clusters_2level}. We divide the set of $N$ data points
into $q_0$ superclusters, and each of these into $q_1$ clusters (in
the Fig. $q_0=q_1=3$). Similarities are drawn independently for
each pair of points. If points are in the same cluster, we use a
Gaussian $N_{\alpha_1,1}(S)$ of mean $\alpha_1$ and variance 1, if
they are in the same supercluster but not in the same cluster, we use
$N_{\alpha_0,1}(S)$, and for all pairs coming from different
superclusters we draw similarities from $N_{0,1}(S)$. The means
fulfill $0<\alpha_0<\alpha_2$.

\begin{figure}[htb]
%\vspace{1cm}
\begin{center}
\includegraphics[width=\columnwidth]{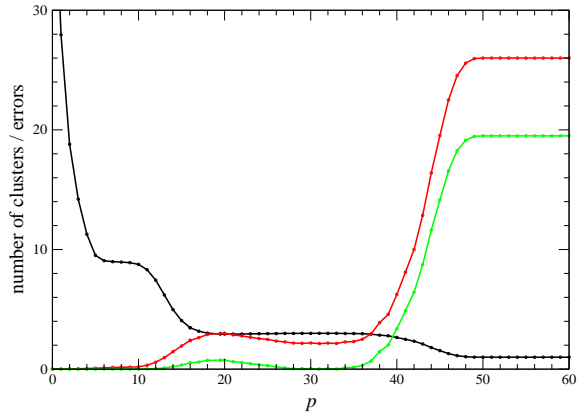}
\end{center}
\caption{SCAP for a systems with two hierarchical levels of
clustering, for $N=180,\ q_0=q_1=3,\ \alpha_0=3,\ \alpha_1=6$,
averages are performed over 2000 samples. The black line shows the
cluster number, two clear plateaus at 9 clusters resp. 3
super-clusters are observed. The red line gives the number of data
points selecting an exemplar in a different cluster, the green line
even in a different super-cluster. Both quantities are divided by 6 to
put them on the same scale as the cluster number. }
\label{fig:scap_2level}
\end{figure}

Fig.~\ref{fig:scap_2level} shows the findings for $N=180,\ q_0=q_1=3,\
\alpha_0=3,\ \alpha_1=6$. We clearly see that SCAP is able to uncover
both cluster levels, pronounced plateaus appear at 3 and 9
clusters. The plot also shows two different error measures: The number
of points which choose an exemplar which is not in the same cluster
(red line in the figure), and the number of points choosing even an
exemplar in a different supercluster (green line). As long as we have
more than 9 clusters, there are very few of both error types
(increasing $\alpha_0$ further decreases this number).  Once we force
clusters at the finest level to merge, the first type of error starts
to grow. The second grows if we observe some merging of
superclusters, i.e. if the cluster number found by SCAP is around or
below 3. Note the little bump in the errors at the beginning of the
three-cluster plateau: There even some links between different
superclusters appear. In fact, in this region the algorithm does not
converge in messages in many cases, leading to many errors. In the
middle of the plateau, however, convergence is much more stable and
error rates are small.

To summarize this section, SCAP is able to infer the cluster structure
of artificial data, even if the latter are organized in a
hierarchical way. Results are very robust and show less errors than
the AP with its hard constraints.

\section{Extension to semi-supervised clustering}
\label{sec:semi}

In case labels are
provided for some data points, they can be exploited to enhance the
algorithmic performance. We propose the following way: Identically
labeled data are collected in {\it macro-nodes}, one for each
label. Since macro-nodes are labeled, they do not need an exemplar,
but they may serve as exemplars for other data. If there are $N$ {\it
unlabeled} points and $m$ known labels, the exemplar mapping thus gets
generalized to ${\bf c}: \{1,...,N\} \mapsto \{1,...,N,N+1,...,N+m\}$
where indexes $N+1,...,N+m$ correspond to macro-nodes. We define the
similarity of an arbitrary unlabeled point to a macro-node as the
maximum of similarities between the point and all elements of the
macro-node
\footnote{Other choices, such as taking the average or center of mass
distance, have been tried, but lead to worse results.}. The Hamiltonian
now becomes:
\begin{equation}
{\cal H}_2[{\bf c}] = -\sum_{\mu=1}^{N} S(\mu,c_\mu)
+p_1\sum_{\mu=1}^N\chi_{\mu}[{\bf c}]+p_2\sum_{\nu=N+1}^{N+m}
\chi_{\nu}[{\bf c}]
\label{eq:cost2}
\end{equation}
Note that neither the sizes of the training set nor of the macro-nodes
appear explicitly. They are implicitly present via the determination
of the similarities between data and macro-nodes. In principle, we can
choose different values of $p_1$ and $p_2$, more precisely $p_1>p_2$,
to reduce the cost of choosing macro-nodes as exemplars as compared to
normal data points. However, this usually forces data to choose the
closest macro-node instead of making a collective choice using the
geometrical information contained in the data set. We found
$p_1=p_2=p$ to work best.

\begin{figure}[htb]
\centerline{\includegraphics[width=\columnwidth]{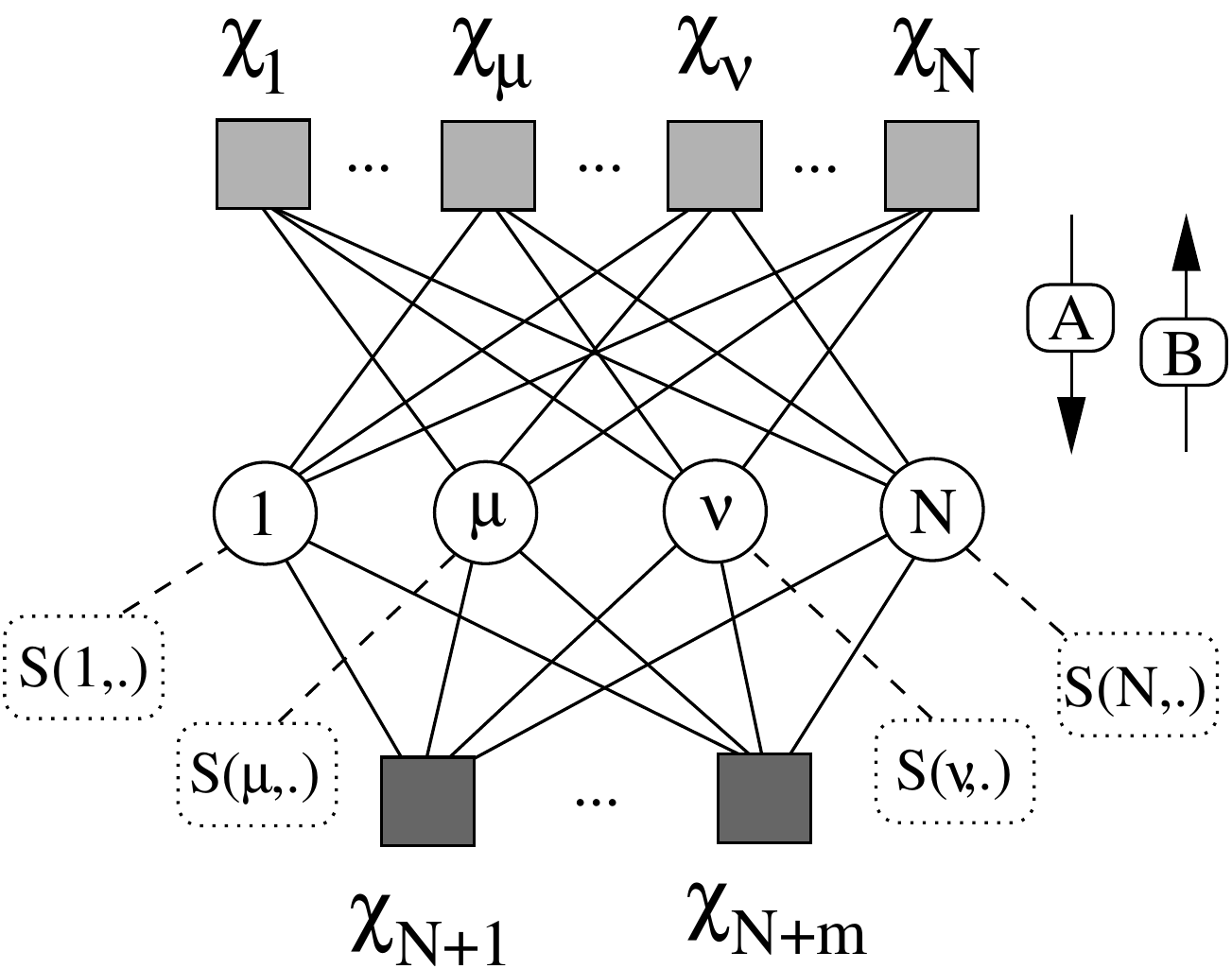}}
\caption{Factor graph and message direction: Circles (variable nodes)
are unlabeled data points, squares (factor nodes) constraints due to
unlabeled (light) and macro-nodes (dark).  Similarities act as
$(N+m-1)$-dimensional external fields on the unlabeled data
points. Messages are exchanged between all connected pairs of data
points and constraints.}
\label{AP-FG}
\end{figure} 
Compared to Fig.~\ref{fig:factor_graph_scap}, the factor graph becomes
slightly more complicated. As is shown in Fig.~\ref{AP-FG}, $m$ new
factor nodes are added to the graph representing the constraints constituted
by the macro-nodes. This modification allows, however, to follow exactly
the same route from the Hamiltonian to the final SCAP equations:
\begin{eqnarray}
\label{eq:semiSCAP1}
a_{\mu\to\nu} &=& {\rm min} [ 0,\, -p
+ \sum_{\lambda\neq\mu,\nu} {\rm max}
(0,\, r_{\lambda\to\mu}) ]  \\
r_{\nu\to\mu} &=&
S(\nu,\mu) -  {\rm max}_{\lambda\neq\mu,\nu}
\left[ S(\nu,\lambda) + a_{\lambda\to\nu} \right] \nonumber
\end{eqnarray}
Remember that $\mu\in\{1,...,N\}$ corresponds to the unlabeled data
points, whereas $\nu\in\{1,...,N+m\}$ enumerates the constraints and
thus the possible exemplars. At infinite $\beta$, the exemplar choice 
becomes polarized to one solution (for non-degenerate similarities) and reads
\begin{equation}
\label{eq:semiSCAP2}
c_\nu^\star = {\rm argmax}_{\mu\in\{1,...,N+m\},\mu\neq\nu} 
\left[  S(\nu,\mu)+a_{\mu\to\nu} \right]\ .
\end{equation}
Compared to Eqs.~(\ref{eq:SCAP1},\ref{eq:SCAP2}) only the number of
constraints becomes modified. The introduction of macro-nodes actually allows
for a very elegant generalization of SCAP from the unsupervised to the 
semi-supervised case.

\subsection{Artificial data}

 To test the performance of
unsupervised vs. semi-supervised SCAP, we turned first to
some artificial cases.

\begin{figure}[htb]
%\vspace{1cm}
\begin{center}
\includegraphics[width=\columnwidth]{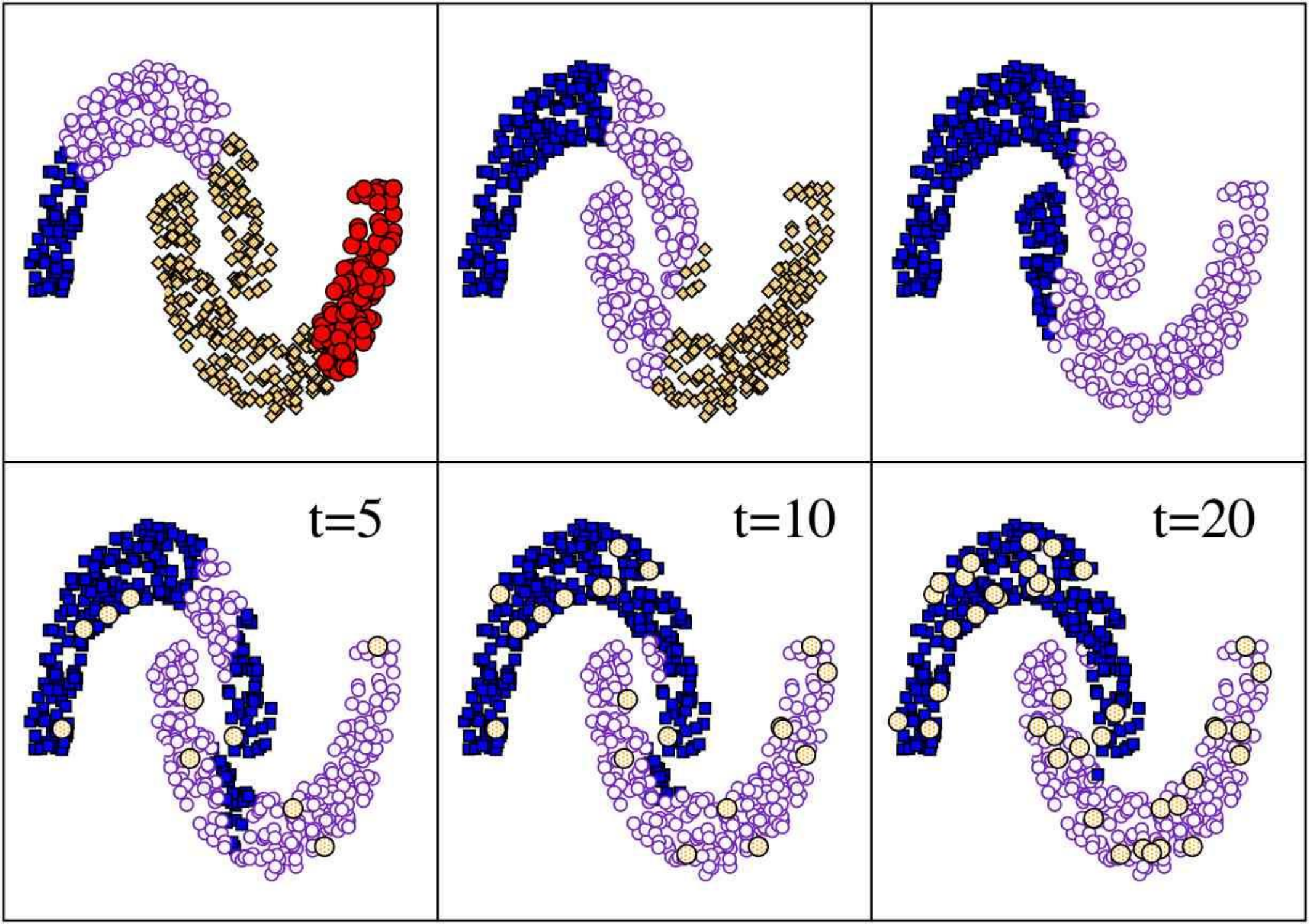}
\end{center}
\caption{Upper row: 3 best clusterings seen by unsupervised
SCAP. $N=600$, $300$ in each cluster.
Lower row: same data set with $t$ trainers (larger circles)
for each cluster.}
\label{small1}
\end{figure}

{\it Data set 1:} We randomly selected points in two dimensions clustered in
a way clearly visible to human eye (Fig.~\ref{small1}). The similarity between
data points is measured by the negative Euclidean distance. The clusters
are so close that the distance between points on the borders of two
clusters is sometimes comparable to the distance between points inside
one single cluster. This makes the clustering by unsupervised methods
harder.  For example, look at Fig.~\ref{small1}, upper row: In this
case, the best unsupervised SCAP clustering makes a significant
fraction of errors, and does not recognize the two clusters. The best results 
with unsupervised SCAP are actually obtained when we allow it to divide the 
data into four clusters.

\begin{figure}[htb]
%\vspace{1cm}
\begin{center}
\includegraphics[width=\columnwidth]{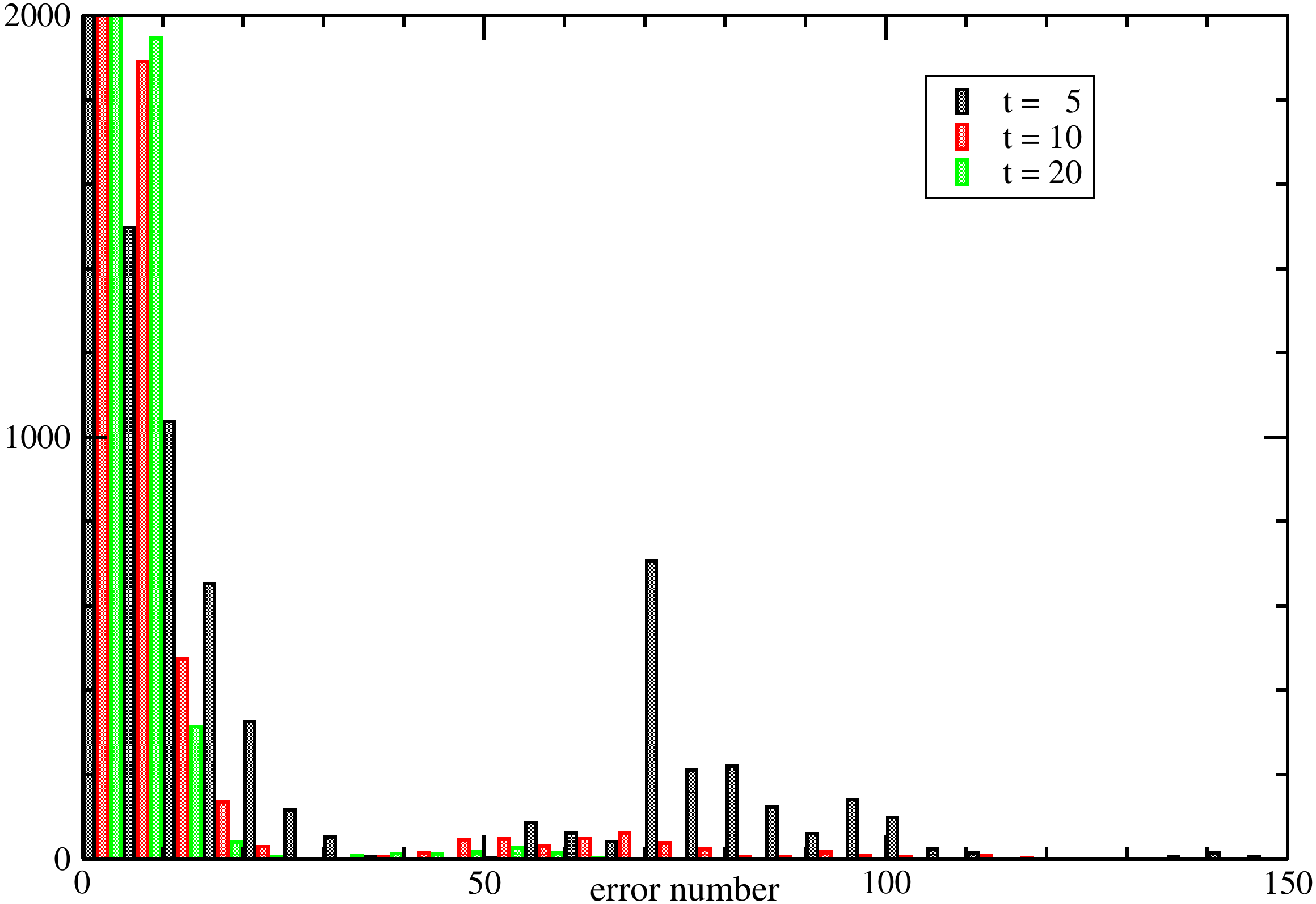}
\end{center}
\caption{Histogram of the number of errors for 10 000 random choices
of $t=5,10,20$ labeled data points. For better visibility, bars are
reduced in width and shifted relative to each other for different 
training-set sizes (bin size 5).}
\label{hist}
\end{figure}

On the other hand semi-supervised SCAP recognizes two clusters very
fast.  When we introduce some labeled points, we find a significant
improvement of the output, cf.~Fig.~\ref{small1}, lower row. Already
as few as 5 labeled points per cluster increase the performance
substantially. Larger training sets lead typically to less errors. In
the semi-supervised SCAP, clustering is very stable and does not
change when we increase $p$.

In Fig.~\ref{small1} we show the clusters for one random choice of
labeled set.  In general one can argue that the clustering would change
with the way the labeled set is distributed inside a cluster. In Fig.
\ref{hist} we show a histogram for 10000 random selections of the
training set, for training set size $t=5,\ 10,\ 20$. We observe that
a majority of clusterings found makes only few errors (the
peak for less then 5 errors is cut in height for better visibility), but a
small number of samples lead to a substantial error number. These
samples are found to have labeled exemplars which are concentrated in
regions mostly far form the regions where clusters are close, so a
relatively large part of these regions is assigned erroneously to the
wrong label. The probability of occurrence of such unfavorable situations
goes down exponentially with the size of the training data set.

\begin{figure}[htb]
%\vspace{1cm}
\begin{center}
\includegraphics[width=\columnwidth]{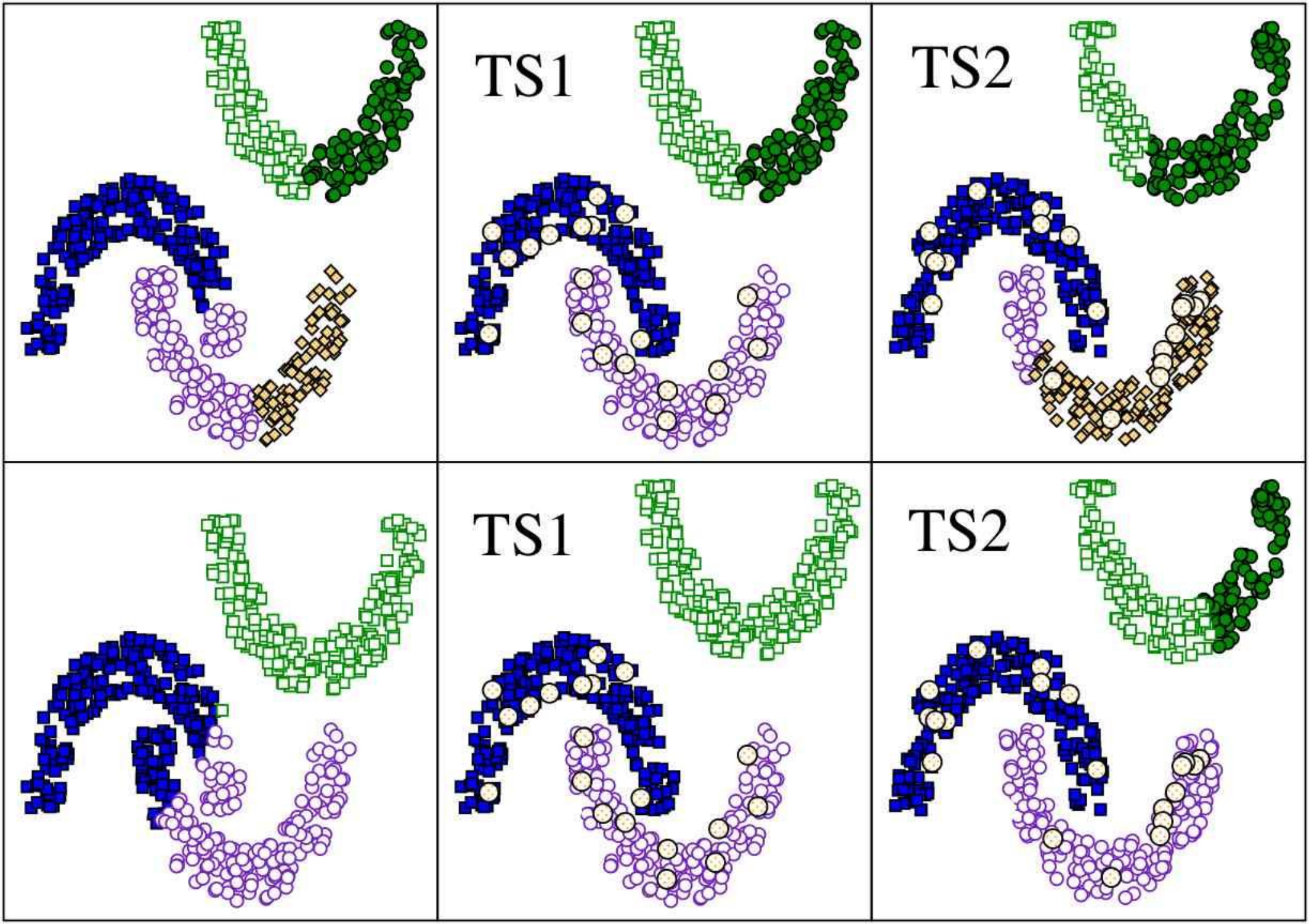}
\end{center}
\caption{Upper row: $N=600$, with 200 data points in each cluster. In all three
cases we choose $p=0.5$.  In the semi-supervised case $t=10$ each of the lower 
clusters; $t=0$ for the upper one.  The two semi-supervised 
results are for different training sets (TS1 and TS2).
Lower row: same data with $p=1$.}
\label{small4}
\end{figure}

\begin{figure}[htb]
%\vspace{1cm}
\begin{center}
\includegraphics[width=\columnwidth]{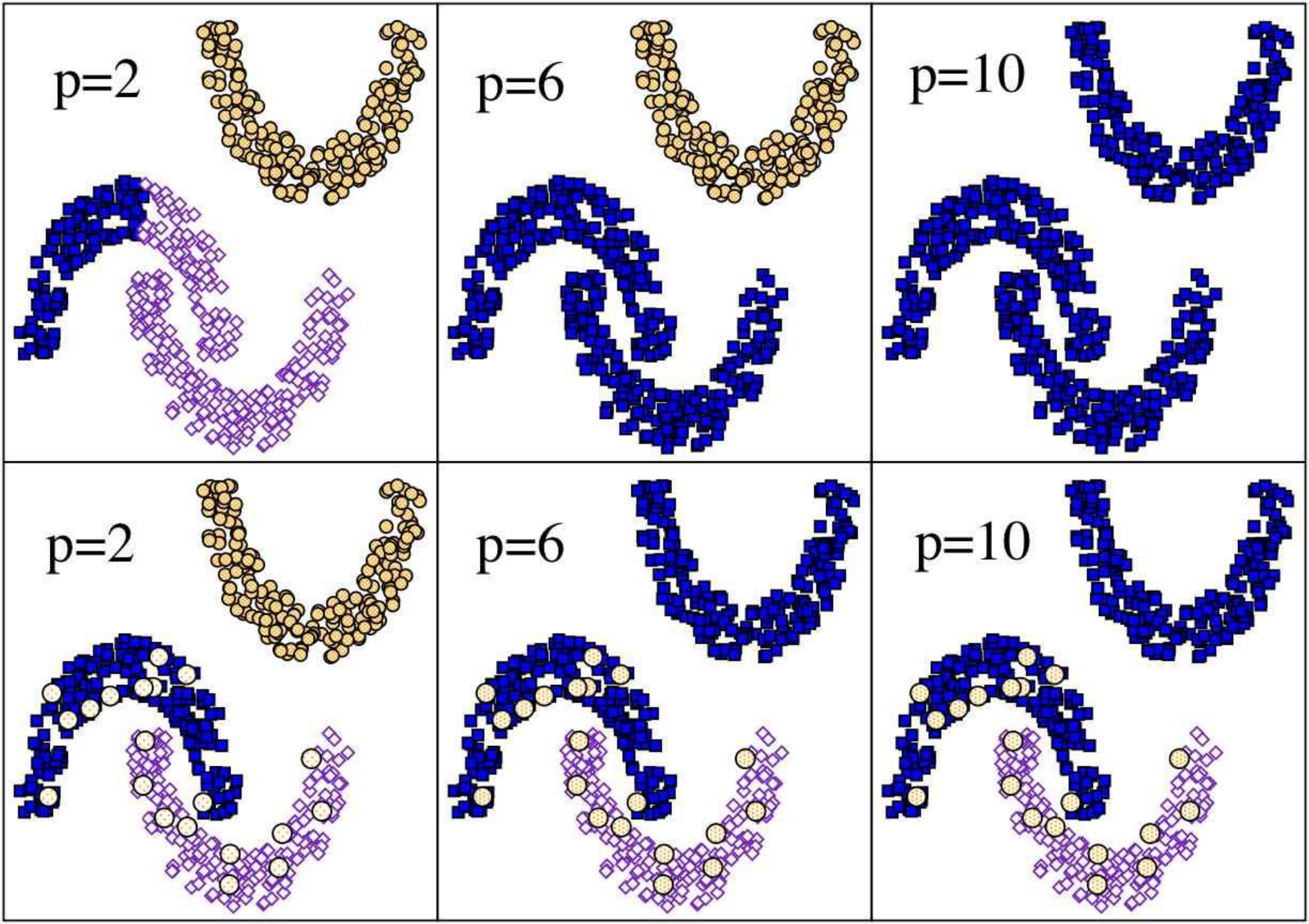}
\end{center}
\caption{$N=600$, with 200 data points in each cluster. First and second
rows contains clustering for unsupervised and semi-supervised
learning for $p=2,6,10$ (left to right). Semi-supervised: 10 trainers each for
lower clusters, 0 for the upper one. One can see how increasing $p$ 
leads to an artificial merging of the labeled clusters ($p=6$). However,
in the Semi-supervised case a stable region of p arises where the third 
cluster is well discerned while the labeled ones are still naturally separated.}
\label{small5}
\end{figure}

{\it Data set 2:} With partial labeling, there are often cases where no
information is available on some of the classes. Semi-supervised SCAP
is able to deal with this situation because it can output clusters
without macro-nodes, i.e. clusters without reference to any of the
trainers' labels.  As an example, we add to the artificial data set a
third cluster of similar size and shape, without adding any new
trainer.  As shown in Fig.~\ref{small4} and
\ref{small5}, the algorithm detects correctly both the labeled and the
unlabeled clusters for a wide range of parameters.

\subsection{Iris data}

This is a classic data set used as a bench mark for testing clustering algorithm \cite{IRIS}. 
The data consist of measurements of sepal length, sepal width, petal length and petal width,
performed for 150 flowers, chosen from three species of the flower Iris.
Unsupervised SCAP already works well making only 9 errors. Introducing $t$ trainers per class, 
the error number further decreases as shown in table~\ref{iris_data_set}.
			    
\begin{table}[htb]
\begin{center}
\begin{tabular}{||c|c|c|c|c||}
\hline
t & 3 & 4-10 & 15-30 & 40\\ 
\hline
errors & 7 & 6 & 2 & 1 \\
\hline
\end{tabular}
\end{center}
\caption{Errors in labeling Iris data, in dependence on the number $t$ of labeled data points.}
\label{iris_data_set}
\end{table}

We also performed semi-supervised clustering
where we provided labels for only two out of the three data sets.
Depending on the number and distribution of labeled points the
algorithm produced 5-9 errors. Semi-supervised SCAP worked better when
we provided information on the clusters corresponding to {\sl
versicolor} and {\sl virginica} species. This is not surprising as
these two are known to be closer to each other than to {\sl setosa},
whose points set is well discerned even in the unsupervised case.

\section{Summary and outlook}
\label{sec:conclusion}

In this paper, a further simplification of soft-constraint affinity
propagation, a message-passing algorithm for data clustering, was proposed.
We have presented a detailed derivation, and have discussed time-
and memory-efficient implementations. The latter are important in particular
for the clustering of huge data sets of more than $10^4$ data points, an
example would be gene which shall be clustered according to their
expression profiles in genome-wide micro-array experiments. Using artificial 
data we have shown that SCAP can be applied successfully to hierarchical
cluster structures, a model parameter (the penalty $p$ for exemplars)
allows to tune the clustering to different resolution scales.
The algorithm is computationally very efficient since it involves updating  
${\cal O}(N^2)$ messages, and it converges very fast.

SCAP can be extended to semi-supervised clustering in a straightforward way. 
Semi-supervised SCAP shares the algorithmic simplicity and stability properties 
of its unsupervised counter part, and can be seen as a natural extension. The
algorithm allows to assign labels to previously unlabeled data, or to
identify additional classes of unlabeled data. This generalization allows to
cluster data even in situations where cluster shapes are involved, and some
additional information is needed to distinguish different clusters.

In its present version, SCAP does not yet fully exploit the
information contained in the messages, only the maximal excess
similarity is used to determine the most probable exemplar. In the
case where labels are not exclusive, one can also use the information
provided by the second, third etc. best exemplar. This could be interesting
in particular in cases, where similarity information is sparse, a popular
example being the community search in complex networks.

In a future work we will explore these directions in parallel to a
theoretical analysis of the algorithmic performance on artificial
data, which will provide a profound understanding of the strength and
also the limitations of (semi-supervised) SCAP. 

{\sl Acknowledgments}:
We acknowledge useful discussions with Alfredo Braunstein and Andrea Pagnani. 
The work of S. and M.W. is supported by the EC via the STREP GENNETEC 
(``Genetic networks: emergence  and complexity'').

%\begin{figure}[htb]
%\vspace{0.2cm}
%\begin{center}
%\includegraphics[width=0.5\columnwidth]{gene_regulation}
%\end{center}
%\caption{}
%\label{fig:gene_regulation}
%\end{figure}

\end{document}